\documentclass{article}
\usepackage[utf8]{inputenc}
\usepackage[a4paper, total={6in, 8in}]{geometry}
\usepackage{authblk}
\usepackage{url}
\usepackage{biblatex}
\usepackage{amsmath}

\usepackage[export]{adjustbox}
\usepackage{comment}
\usepackage{float}
\usepackage[linesnumbered,ruled,vlined]{algorithm2e}
\usepackage{dirtytalk}
\usepackage{verbatim}
\usepackage{amssymb}
\usepackage{soul}
\usepackage{amsmath}
\usepackage{dirtytalk}
\usepackage{multirow}
\usepackage{enumitem}
\usepackage{amsmath}
\usepackage{amsthm}

\newtheorem{theorem}{Theorem}
\newtheorem{definition}{Definition}

\newtheorem{proposition}{Proposition}

\title{\textbf{Civic Crowdfunding for Agents with Negative Valuations and Agents with Asymmetric Beliefs}}

\author[1]{Sankarshan Damle\thanks{sankarshan.damle@research.iiit.ac.in}}
\author[1]{Moin Hussain Moti\thanks{moin.moti@research.iiit.ac.in}}
\author[2]{Praphul Chandra\thanks{praphulcs@koinearth.com}}
\author[1]{Sujit Gujar\thanks{sujit.gujar@iiit.ac.in}}
\affil[1]{Machine Learning Lab, International Institute of Information Technology, Hyderabad}
\affil[2]{KoineArth Inc., Bangalore}

\date{}

\begin{document}

\maketitle

\begin{abstract}  
In the last decade, civic crowdfunding has proved to be effective in generating funds for the provision of public projects. However, the existing literature deals only with citizen's with positive valuation and symmetric belief towards the project's provision.  In this work, we present novel mechanisms which break these two barriers, i.e., mechanisms which incorporate negative valuation and asymmetric belief, independently. For negative valuation, we present a methodology for converting existing mechanisms to mechanisms that incorporate agents with negative valuations. Particularly, we adapt existing PPR and PPS mechanisms, to present novel PPRN and PPSN mechanisms which incentivize strategic agents to contribute to the project based on their true preference. With respect to asymmetric belief, we propose a reward scheme Belief Based Reward (BBR) based on Robust Bayesian Truth Serum mechanism. With BBR, we propose a general mechanism for civic crowdfunding which incorporates asymmetric agents. We leverage PPR and PPS, to present PPRx and PPSx. We prove that in PPRx and PPSx, agents with greater belief towards the project's provision contribute more than agents with lesser belief. Further, we also show that contributions are such that the project is provisioned at equilibrium.
\end{abstract}

\section{Introduction}
\emph{Crowdfunding} is a process of raising funds from a large pool of interested agents and is an active area of research \cite{Alaei:2016:DMC:2940716.2940777,Itai2017,REPPS2017,strausz2017theory,Shen2018}. The process, when applied for the provision of public projects, is called \emph{civic crowdfunding}. In the last decade, civic crowdfunding has grown to be instrumental in providing a platform through which citizens can collectively finance social initiatives such as libraries, public parks, etc. 

In the standard approach for civic crowdfunding, the social planner uses the voluntary contribution mechanism with a provision point, the \emph{provision point mechanism} (\cite{Bagnoli1989}). The social planner sets up a target amount, referred to as the provision point, to be raised. If the contributions by the agents exceed the provision point, social planner provisions the project; otherwise, returns the contributions. The mechanism, however, has been shown to have several inefficient equilibria \cite{Bagnoli1989,Brubaker1975,Schmidtz1991}. 

Provision Point mechanism with Refund bonus (PPR) by \cite{PPR2014} introduces an additional \emph{refund bonus} to be paid to all the contributing agents (along with their contribution) in case the project is not provisioned. Chandra et. al. \cite{PPS2016} showed that in \emph{sequential setting}, wherein the history of contributions is known to the agents, PPR collapses to a simultaneous move game, among the contributing agents. Towards this, they proposed Provision Point mechanism with Securities (PPS) with refunds based on complex prediction markets, and showed that it induces a sequential game, in which the project is provisioned at equilibrium. Thus in this paper, for a sequential game, we focus on PPS while focusing on PPR for a simultaneous game. To the best of our knowledge, there are no other refund bonus mechanisms for civic crowdfunding.

Note that in all these mechanisms only those agents with a positive valuation towards the project contribute to its provision. However, several agents may \emph{prefer} the project to not be provisioned, i.e., their valuation may be negative for the project getting provisioned. For instance, consider the construction of a garbage dump yard in a locality. While the project may be welcomed by a number of agents, a certain set of agents may wish to relocate the project from its current location to another. In other words, these agents may not \emph{prefer} the construction of the dump yard -- in the locality proposed. In such a scenario, the construction of the dump yard (as well as the locality in which it is constructed) must depend on the majority’s opinion of it. The civic crowdfunding literature does not address such negative valuation. If addressed, civic crowdfunding with agents with negative valuation can provide a natural way for \emph{preference aggregation}.

We define an agent's \emph{information structure} as consisting of its valuation and its belief towards the project's provision. Based on their valuation, we categorize these agents as follows: \emph{positive (negative) agents} i.e., agents with positive (negative) valuation or \emph{positive (negative) preference} towards the project's provision. The mechanisms, mentioned above, also assume that apart from knowing the history of contributions, agents do not have any information regarding the provision of the project, i.e., every agent's belief is \emph{symmetric} towards the project's provision. 

Motivated to break these barriers on an agent's information structure in existing literature for civic crowdfunding, in this paper, we address these two limitations by (i) handling symmetric agents with negative preference  and (ii) handling positive agents with \emph{asymmetric} belief towards the project's provision, independently. Relaxing both the assumptions in one mechanism is still illusive. 

To incorporate civic crowdfunding for agents with negative valuation, we require mechanisms that integrate negative agents. For this, we set up two parallel markets, with two different targets -- one for the provision, i.e., \emph{provision point} and one against the provision, i.e., \emph{rejection point}, for the project. The project is provisioned (not provisioned) if the provision (rejection) point is reached first.  A strategic agent may choose to contribute in a market, against its preference. Thus, the challenge in such a setting remains to ingeniously design a refund scheme such that the agents are incentivized to contribute based on their preference. For this, we propose a methodology through which existing mechanisms for positive preferences also allow for agents with negative preferences, such that agents contribute to the market based on their true preference. In particular, we adapt existing PPR and PPS mechanisms to design PPRN and PPSN mechanisms. We prove that in these mechanisms at equilibrium, either the provision point or the rejection point holds. 

Further, designing mechanisms for civic crowdfunding for agents with \emph{asymmetric} beliefs is not trivial. For instance, a rational agent with significant belief towards the project's provision may choose to free-ride, as it believes that the project will be provisioned regardless of its contribution. Such asymmetric agents need to be \emph{further} incentivized to contribute towards the project's provision. For this, we propose a novel reward scheme \emph{Belief Based Reward} (BBR) that rewards an agent based on their belief towards the project's provision. We deploy a \emph{peer prediction} mechanism for information aggregation of each agent's belief. With BBR, we propose a novel class of mechanisms for civic crowdfunding which incentivizes agents with asymmetric beliefs to contribute towards the provision, such that the project is provisioned at equilibrium.

\section{Preliminaries}
In this section, we present the required preliminaries. We begin by defining our crowdfunding model.
\subsection{Model}

	For the civic crowdfunding of public project $PP$, the Project Maker (PM), sets up a market for its provision. The PM announces a provision point $h^0$ as the target to be reached until a deadline $T$.  Let $\mathbb{A}=\{1,\dots,n\}$ be the set of all agents wherein each Agent $i$ has valuation $\theta_i\geq 0$ if $PP$ gets provisioned. The agents contribute $x=(x_1,\dots,x_n)$ to the crowdfunding mechanism. Let $\vartheta = \sum_{i=1}^{i=n} \theta_i$; the total valuation of all the agents and $\mathcal{X}=\sum_{i=1}^{i=n} x_i$; the sum of the contributions. A project is provisioned if $\mathcal{X}\geq h^0$  at the end of deadline $T$ and not provisioned otherwise. Such mechanisms are referred to as \emph{provision point} mechanisms. Note that $\theta_i$s are private to the agents and provision point mechanisms are indirect mechanisms to aggregate these. This setup induces a game amongst the agents.     
	

	Let $\sigma = (\sigma_1,\dots,\sigma_n)$ be the vector of strategy profile of every agent. Agent $i$'s strategy consists of its contribution $x_i$ towards the project's provision along with other mechanism dependent parameters. We use the subscript $-i$ to represent vectors without Agent $i$. The utility for Agent $i$ with valuation $\theta_i$ for the project, when all the agents play the strategy profile $\sigma$ is $u_i(\sigma;\theta_i)$.  
	
	We now define some important game-theoretic definitions necessary for the analysis of the mechanisms presented in this paper.

    	\begin{definition}[Nash Equilibrium (NE)]
        	A strategy profile $\sigma^* = (\sigma_1^*,\dots,\sigma_n^*)$ is said to be a Nash equilibrium (NE) if for every Agent $i$, it maximizes the utility $u_i(\sigma^*;\theta_i)$ i.e., $\forall i \in \mathbb{A}$, $$u_i(\sigma_i^*,\sigma^*_{-i};\theta_i) \geq u_i(\sigma_i,\sigma^*_{-i};\theta_i) \ \forall \sigma_i, \forall \theta_i.$$
        	\end{definition}
    
    For crowdfunding in sequential setting, i.e., when the contributing agents arrive over time, the strategy profile of every Agent $i$ also constitutes the time $t_i$ at which it contributes to the mechanism. Let, $a_i$ be the time at which Agent $i$ arrives to the mechanism. Further, let $h_t^0$ denote the amount remaining for the project to be provisioned at time $t$. With this, we define,  
    
	\begin{definition}[Sub-game Perfect Equilibrium (SPE)] A strategy profile $\sigma^* = (\sigma_1^*,\dots,\sigma_n^*)$ is said to be a sub-game perfect equilibrium if for every Agent $i$, it maximizes the utility $u_i(\sigma^*_i,\sigma^*_{-i|H^{a_i}};\theta_i)$ i.e., $\forall i \in \mathbb{A}$,
            $$u_i(\sigma^*_i,\sigma^*_{-i|H^{a_i}};\theta_i) \geq u_i(\sigma_i,\sigma^*_{-i|H^{a_i}};\theta_i) \ \forall\sigma_i, \forall H^{t}, \forall \theta_i.$$ 
            \end{definition}
    
    Here, $H^t$ is the history of the game till time $t$, constituting the agents' arrivals and their contributions and $\sigma^*_{-i|H^{a_i}}$ indicates that the players who arrive after $a_i$ follow the strategy specified by $\sigma^*_{-i}$. 

 In the next subsection we describe existing mechanisms in the literature for civic crowdfunding.

    \subsection{Provision Point Mechanisms}
In this paper, we focus on the class of mechanisms for civic crowdfunding which require the project to aggregate a minimum level (Provision Point of the project) of funding before the PM can claim it. Several provision point mechanisms for civic crowdfunding have been proposed \cite{Brubaker1975,Groves1977,Bagnoli1989,Schmidtz1991,Tabarrok1998,Chen2008,REPPS2017}; but our work focuses on PPR \cite{PPR2014} and PPS \cite{PPS2016} mechanisms. 
	\subsubsection{Provision Point Mechanism with Refund (PPR)}
   To counter the problem of free-riding, PPR offers a refund bonus to the agents in case the project does not get provisioned. The refund bonus scheme is directly proportional to agent's contribution. Let $\mathcal{I}_X$ be an indicator random variable which takes the value $1$ if $X$ is true and $0$ otherwise. Then the utility structure of PPR, for each agent $i \in \mathbb{A}$ is given as, 
\begin{equation*}
	u_i = \mathcal{I}_{\mathcal{X} \geq h^0} \cdot (\theta_i - x_i) + \mathcal{I}_{\mathcal{X} < h^0} \cdot \left(\frac{x_i}{\mathcal{X}}\right)B,
\end{equation*}
where $B$ is the total budget kept aside by the PM, and is distributed to the agents who contributed along with their contribution, as a refund, in case the project is not provisioned.

The refund bonus in PPR is independent of time of contribution and therefore all agents delay their contributions as close to the deadline as possible and wait to free-ride till the end. Such strategies lead to the project not getting provisioned in practice and are therefore undesirable.
    \subsubsection{Provision Point Mechanism with Securities (PPS)}
PPS addresses the shortcomings of PPR by offering agents refunds based on the time of their contribution. An early contributor is paid higher refund than a late contributor for the same contribution. The utility structure of PPS, for each agent $i \in \mathbb{A}$ is given as,
\begin{equation*}
	u_i = \mathcal{I}_{\mathcal{X} \geq h^0} \cdot (\theta_i - x_i) + \mathcal{I}_{\mathcal{X} < h^0} \cdot (r^{t_i}_i - x_i),
\end{equation*}
where, $t_i ~ \mbox{and} ~ r_i^{t_i}$ are Agent $i$'s time of contribution and the number of securities allocated to it, respectively. $r_i^{t_i}$ depends on its contribution $x_i$ at $t_i$, as well as the total number of securities issued in the market at time $t_i$, denoted by $q^{t_i}$, i.e. \cite[Eq. 6]{PPS2016},

    $$r_i^{t_i}=C_0^{-1}\left(x_i+C_0(q^{t_i})\right)-q^{t_i}.$$
    
    Here, $C_0$ is the \emph{cost function} governing the underlying prediction market in PPS obtained from the general cost function $C$, by fixing the number of positive outcome securities. A cost function must satisfy \cite[CONDITIONS 1-4,6]{PPS2016} to be used in PPS. The properties of the cost function, $C_0$, relevant to this paper include:
        \begin{enumerate}
            \item  $r_i^{t_i}$ is an increasing function of  $x_i$, i.e., $\frac{\partial r_i^{t_i}}{x_i}>1$, $\forall \theta<h^0$ \cite[CONDITION-7]{PPS2016}.
            \item $r^{t_i}$ is a decreasing function of $t_i$ \cite[Step-2 (Theorem 3)]{PPS2016}.
        \end{enumerate}
    
The existing literature for civic crowdfunding is limited through its assumptions on the information structure of the contributing agents. We define an agent's information structure as consisting of its valuation and its belief towards the project's provision. Based on their valuation, we categorize agents as follows: positive (negative) agent i.e., Agent $i$ with $\theta_i\geq 0$ ($\theta_i<0$) or with positive (negative) preference towards $PP$'s provision. Let $\mathbb{P}$ ($\mathbb{N}$) denote the set of all positive (negative) agents, such that $\mathbb{A}=\mathbb{P}\cup\mathbb{N}$. Further, let $\vartheta^1=\sum_i \theta_i \: \forall i \in \mathbb{P}$ as the total valuation for $PP$ getting provisioned and $\vartheta^2=\sum_i\left(-\theta_i\right) \: \forall i \in \mathbb{N}$ as the total valuation for $PP$ not getting provisioned, i.e., $\vartheta=\vartheta^1 - \vartheta^2$. For preference aggregation, the PM's goal is to determine whether the majority prefers $PP$ to be provisioned or not, i.e., whether or not $\vartheta \geq 0$.

   We also consider agents with asymmetric beliefs towards $PP$'s provision i.e., agents may believe that the project may be provisioned with probability (belief) $1/2\pm \epsilon$ or may not be provisioned with probability $1/2\mp\epsilon$ for some $\epsilon \geq 0$. Let $k^1_i = (1/2 + \epsilon_i)$ and $k_i^2 = (1/2 - \epsilon_i)$ for some $\epsilon_i \geq 0$ such that $k_i^1 + k_i^2 = 1$, $\forall i \in \mathbb{A}$. Let $\mathbb{A}^+$ ($\mathbb{A}^-$) be the set of agents which believe that $PP$ will (will not) be provisioned i.e., every agent $i \in \mathbb{A}^+$ ($i \in \mathbb{A}^-$) has belief $k^1_i$ ($k^2_i$) that $PP$ will be provisioned, such that $\mathbb{A}=\mathbb{A}^+\cup\mathbb{A}^-$.
    
In this paper, we require each agent to truthfully elicit its belief regarding the provision of the public project. Since an agent's opinion (belief) is its private information, we look for mechanisms which incentivize it to elicit its true opinion. In mechanism design theory, such mechanisms are called \emph{incentive compatible} (IC). Further, these mechanisms must also be \emph{individually rational} (ID) i.e., each Agent $i$ must have non-negative utility. Towards this, we make use of peer prediction mechanisms.
    
    \subsection{Peer Prediction Mechanisms}
    Peer prediction mechanisms (PPM) allow for elicitation and aggregation of subjective opinions from a set of agents. These are generally deployed in situations where there is no method of verifying an agent's honesty (of their opinion) or their ability. In the literature, there are a number of existing peer prediction mechanisms \cite{Miller2005,Jurca2007,Lambert2008,Dasgupta2013,Radanovic2013}. 
    
    For illustrative purposes, in this paper, we focus on Robust Bayesian Truth Serum (RBTS) mechanism (\cite{RBTS2014}). 
		\subsubsection{Robust Bayesian Truth Serum}
        While RBTS mechanism properties hold for an arbitrary number of signals, we present the binary version of the mechanism, since we are interested in the elicitation of an agent's belief towards the provision of the project. In RBTS, each agent is required to submit, from \cite{RBTS2014}: 
	\begin{enumerate}
	\item \textbf{Information Report:} Let $f_i = \{0,1\}$ be Agent $i$'s reported
signal.
    \item \textbf{Prediction Report:} Let $g_i \in [0, 1]$ be Agent $i$'s report
about the frequency of high signals among the citizens.
	\end{enumerate}

Based on information report and prediction report, RBTS assigns a score to each agent. The mechanism, for each Agent $i$, selects a \emph{reference} agent $j=i+1\mbox{~(modulo~}n)$ and a \emph{peer} agent $k=i+2\mbox{~(modulo~}n)$ and calculates,
		\[
 	  g_i' = 
 	  \begin{cases}
		g_j + \delta & \mbox{~if~} f_i = 1 \\ 
        g_j - \delta & \mbox{~if~} f_i = 0 \\ 
	  \end{cases}
		\]
where $\delta =\mbox{min}(g_j,1-g_j)$. Then, the RBTS score for Agent $i$ is given by, 	\begin{equation}
	 \label{eq:RBTS}
     RBTS_i = \underbrace{G_m(g_i',f_k)}_\textrm{
information score} + \underbrace{G_m(g_i,f_k)}_\textrm{
prediction score}.
     \end{equation}

$G_m(\cdot)$ is the binary quadratic scoring rule \cite{selten1998axiomatic} normalized to give scores between 0 and 1 and is a strictly proper scoring rule.

 RBTS mechanism is IC and ex-post ID (\cite[Theorem 10]{RBTS2014}) for the elicitation of binary information for all $n \geq 3$,  without relying on knowledge of the \emph{common prior}. Note that many other peer prediction mechanisms are IC only when $n\rightarrow \infty$ and hence we chose RBTS. 

\section{Civic Crowdfunding for Agents With Negative Valuations}
\label{section::preference}
In this section, we introduce a methodology through which civic crowdfunding mechanisms can incorporate \emph{symmetric} agents with negative preference (valuation) towards the public project's provision. For this, the PM sets up two separate markets, i.e., one for the provision and one against the provision of the project. Thus, agents now have a greater scope for manipulation. In such a setting, a strategic agent may choose to contribute in a market, against its preference, if its expected utility for contributing in that market is more than if it contributes in the market based on its preference. Therefore, to incorporate agents with negative preference, we must ingeniously construct the refund scheme in a way so that the agents are incentivized to contribute in the market based on their true preference. 

To illustrate this methodology, we provide two mechanisms for the same by adopting existing mechanisms in PPR and PPS; as namely, PPRN and PPSN. For these mechanisms, let $h^1$ ($h^2$) be the target for provision (rejection) of $PP$ with $\mathcal{X}^1$ ($\mathcal{X}^2$) as the total funding received towards (against) its provision. Further, let $h^1_{t}$ ($h^2_{t}$) denote the amount remaining for the project to be provisioned (not provisioned) at time $t$. 

We now present \emph{Provision Point Mechanism with Refund for Negative Preference} (PPRN) leveraging PPR. 

\subsection{Provision Point Mechanism with Refunds with Negative Preference (PPRN)}

We now propose a mechanism for civic crowdfunding for agents with negative valuation by leveraging PPR, namely PPRN.
	\subsubsection{Protocol} In PPRN, Agent $i$ not only contributes its contribution $x_i$ but also specifies its preference i.e., whether it wants to contribute towards the project getting provisioned or not getting provisioned.
	Let $s_i \in \{1,2\}, \: \forall i \in \mathbb{A}$ be a private preference variable for Agent $i$, such that $s_i=1,\forall i \in \mathbb{P}$ and $s_i=2,\forall i \in \mathbb{N}$. Further, let Agent $i$'s reported preference be $\Tilde{s}_i$. PM adds Agent $i$'s contribution towards the project's provision or rejection based on its reported preference $\Tilde{s}_i$. The project is provisioned or not based on whichever target is first reached.
       
	\subsubsection{Agent Utility} The utility for Agent $i\in \mathbb{A}$ with $\Tilde{s}_i = 1$ in PPRN is as follows,
    \begin{equation}\tag{PPRN-U1}
	\label{eqn1::PPRN}
	u_i(\cdot) = I_{\mathcal{X}^1\geq h^1}\cdot (\theta_i - x_i) + I_{\mathcal{X}^1< h^1}\cdot \left(\frac{x_i}{\mathcal{X}^1+\mathcal{X}^2}\right)B
	\end{equation}
  	Similarly, the utility for Agent $i\in \mathbb{A}$ with $\Tilde{s}_i = 2$ in PPRN is as follows,
    \begin{equation}\tag{PPRN-U2}
    \label{eqn2::PPRN}
    u_i(\cdot) = I_{\mathcal{X}^2\geq h^2}\cdot \left(-x_i\right)  + I_{\mathcal{X}^2< h^2}\cdot\left(\theta_i +\left(\frac{x_i}{\mathcal{X}^1+\mathcal{X}^2}\right)B\right)
	\end{equation}

\subsubsection{Equilibrium Analysis} We now provide the equilibrium analysis for PPRN as the following theorem,
	\begin{theorem}
    \label{Theorem::PPRN}
	For PPRN, with the utility as given by Eq. \ref{eqn1::PPRN} and Eq. \ref{eqn2::PPRN} $\forall i \in \mathbb{A}$, which satisfies $0<B < \frac{(h^1+h^2)(\vartheta^1-h^1)}{h^1}$, $0 <B < \frac{(h^1+h^2)(\vartheta^2-h^2)}{h^2}$, $\vartheta^1>h^1$ and $\vartheta^2>h^2,$ a set of strategies $\sigma_i^* = (x_i^*,t_i^*,\Tilde{s}_i)$, 
    \[
 	  \sigma_i^* =
 	  \begin{cases}
      	(0,T,s_i) \quad \mbox{~if~} \exists \ l \mbox{~s.t.~} h^l_{T}=0, \mbox{~else,} \\
       (x_i^*,T,s_i) : x_i^* < \left(\frac{h^1+h^2}{B+h^1+h^2}\right)|\theta_i|
	  \end{cases}
     \]  $\forall l\in\{1,2\}$; form a set of equilibrium strategies $\forall i \in \mathbb{A}$, such that at equilibrium either $\mathcal{X}^1 = h^1 \mbox{~or~} \mathcal{X}^2 = h^2$ holds.
	\end{theorem}
\noindent\textbf{Proof.}  In Step 1, we show that the equilibrium contributions are such that at equilibrium either $\mathcal{X}^1=h^1$ or $\mathcal{X}^2=h^2$ holds. Step 2 shows that each agent delays its contribution till the deadline $T$. We prove that every agent contributes based on its true preference in Step 3. Step 4 calculates the equilibrium contribution of every agent. Finally, in Step 5 we give the conditions for existence of Nash Equilibrium. \\
\noindent\underline{\emph{Step 1}}: As $\vartheta^1> h^1$ and  $\vartheta^2> h^2$, at equilibrium $\mathcal{X}^1<h^1$ and $\mathcal{X}^2<h^2$ cannot hold, as $\exists i\in \mathbb{P}\mbox{~and~}\exists j\in \mathbb{N}$ with $x_i<\theta_i$ and $x_j<|\theta_j|$, at least, that could obtain a higher refund bonus by marginally increasing its contribution because of $B>0$. Likewise, any agent with a positive contribution could gain in utility by marginally decreasing its contribution if $\mathcal{X}^1>h^1$ or $\mathcal{X}^2>h^2$. Thus, at equilibrium, either $\mathcal{X}^1=h^1$ or $\mathcal{X}^2=h^2$ holds. \\
\noindent\underline{\emph{Step 2}}: As the refund in PPRN is independent of time, no agent has any incentive to contribute early. Thus, each agent delays its contribution till the deadline, $T$.\\
\noindent\underline{\emph{Step 3}}: Since every Agent $i$ is symmetric in its belief towards the project's provision, its expected utility is given by $\frac{1}{2}\left(\theta_i-x_i+\frac{x_i}{\mathcal{X}^1+\mathcal{X}^2}B\right)$. Thus, every Agent $i$ has no incentive to deviate from its preference. Thus, $\Tilde{s}_i=s_i,\forall i \in \mathbb{A}$.\\
\noindent\underline{\emph{Step 4}}: 
    	\begin{enumerate}
	\item \emph{For Positive agents:} At equilibrium, the \emph{best response} for an Agent $i\in\mathbb{P}$ is that contribution $x_i^*$ such that its \emph{provisioned utility is not less than not provisioned utility} since it prefers the project to be provisioned and is symmetric in its belief, i.e.,  $\forall i \in \mathbb{P}$,
             	$$\theta_i - x_i^* \geq  \left(\frac{x_i^*}{\mathcal{X}^1+\mathcal{X}^2}\right)B$$
            	\begin{equation*}
            	\Rightarrow x_i^* \leq \left(\frac{(\mathcal{X}^1+\mathcal{X}^2)}{B+\mathcal{X}^1+\mathcal{X}^2}\right)\theta_i < \left(\frac{h^1+h^2}{B+h^1+h^2}\right)\theta_i,
            	\end{equation*} Last inequality follows from the fact that $\mathcal{X}^1+\mathcal{X}^2 < h^1 + h^2$.
    \item \emph{For Negative agents:} The \emph{best response} for an Agent $i \in \mathbb{N}$ is that equilibrium contribution $x_i^*$ such that its \emph{not provisioned utility is not less than provisioned utility} since the agent prefers the project to not be provisioned and is symmetric in its belief, i.e., $\forall i \in \mathbb{N}$,
    	$$|\theta_i| +  \left(\frac{x_i^*}{\mathcal{X}^1+\mathcal{X}^2}\right)B \leq -x_i^*$$
            	\begin{equation*}
              \Rightarrow x_i^* \leq \left(\frac{(\mathcal{X}^1+\mathcal{X}^2)}{B+\mathcal{X}^1+\mathcal{X}^2}\right)|\theta_i| < \left(\frac{h^1+h^2}{B+h^1+h^2}\right)|\theta_i| 
            	\end{equation*}
    	\end{enumerate}	
\noindent\underline{\emph{Step 5}}: Summing up $x_i^*,\forall i \in \mathbb{P}$ gives the condition for existence of Nash Equilibrium as,
		\begin{equation*}
		\label{PPRN:Cond1}
	0<B < \frac{(h^1+h^2)(\vartheta^1-h^1)}{h^1}.
     	\end{equation*}
    Similarly, summing up $x_i^*,\forall i \in \mathbb{N}$,
    	\begin{equation*}
    	\label{PPRN:Cond2}
    0 <	B < \frac{(h^1+h^2)(\vartheta^2-h^2)}{h^2}.     
    	\end{equation*} gives the condition for existence of Nash Equilibrium.   \qed

In the next subsection, we present \emph{Provision Point Mechanism for Securities with Negative Preference} (PPSN) by leveraging PPS.

\subsection{Provision Point Mechanism for Securities with Negative Preference (PPSN)}
We now propose a mechanism for civic crowdfunding for agents with negative valuation by leveraging PPS, namely PPSN.
   \subsubsection{Protocol} In PPSN, we consider a mechanism with two \emph{independent} PPS prediction markets - PPS1 and PPS2. In PPS1, agents contribute for the project to be provisioned (and buy negative securities) while in PPS2, agents contribute for the project to not be provisioned (and buy positive securities). Note that the markets being independent, the prices in both the markets are also independent of the other. Provision point for PPS1 is reached when the total contribution in it reaches $h^1$, and rejection point for PPS2 when the total contribution in it reaches $h^2$. Let, $\mathcal{X}^1$ be the total contribution received by the project in PPS1 and $\mathcal{X}^2$ be the total contribution received by the project in PPS2. The project is provisioned or not based on whichever target is first reached.
   
   Let $s_i \in \{1,2\}$, be a private preference variable for Agent $i$, such that $s_i=1,\forall i \in \mathbb{P}$ and $s_i=2,\forall i \in \mathbb{N}$.

	\subsubsection{Common Refund Scheme}
  An agent may not contribute in the market based on its preference if its expected refund is more in case it contributes in the other market. 
To prevent this, we present a \emph{common refund scheme} that ensures that the agent obtains same refund in spite of which market it chooses to contribute. In this, Agent $i$ contributes $x_i$ in any market based on a refund that depends on the minimum of the issued securities present in both the markets i.e., $Q^{t_i}=\textnormal{min}(q^{t_i}_{PPS1},q^{t_i}_{PPS2})$. Based on this, Agent $i$ is issued securities ($R_i^{t_i}$) for a contribution $x_i$ given by $$R_i^{t_i}=C_0^{-1}(x_i+C_0(Q^{t_i}))-Q^{t_i},$$ from \cite[Eq. 6]{PPS2016}. Thus, Agent $i$'s refund in this scheme is $R_i^{t_i}-x_i$.
    
    However, the number of issued securities only changes for the market in which the agent contributes $x_i$ to, i.e.,
    $$C_0^{-1}(x_i+C_0(q^t_{PPS(s_i)}))-q^t_{PPS(s_i)},$$ will be the change in the total number of issued securities in the market PPS($s_i$).
  
\begin{proposition}
\label{PPSN1:Prop1}
The securities allotted to an agent with total issued securities as $Q^{t}$ is always greater than or equal to those it would have received with securities $q^{t}_{PPS1}$ or $q^{t}_{PPS2}$ for the same contribution and the same cost function $C_0$. 
\end{proposition}  
\noindent\textbf{Proof.} The statement follows directly from the fact that the number of securities allotted, for the same contribution, is a decreasing function of the total issued securities \cite[Step-2 (Theorem 3)]{PPS2016}. \qed
  
\begin{proposition}
\label{PPSN1:Prop2}
The refund given by $R^{t_i}_i-x_i$ for Agent $i$, is a decreasing function with respect to time $t_i$.
\end{proposition}  
\noindent\textbf{Proof.} The securities allotted to Agent $i$, $R_i^{t_i}$, decreases as $Q^{t_i}$ increases (Proposition \ref{PPSN1:Prop1}). Further, since $Q^{t}=\textnormal{min}(q^{t}_{PPS1},q^{t}_{PPS2})$ and $q^{t}_{PPS1} \mbox{~and~}q^{t}_{PPS2}$ are non-decreasing with respect to time $t$; $Q^{t}$ is a non-decreasing function of time. Thus, $R_i^{t_i}-x_i$ for Agent $i$, is a decreasing function with respect to time $t_i$. 
\qed

   Let us call PPS1 as $p_1$, and PPS2 as $p_2$. Thus, $p_{s_i}=p_1,\forall i \in \mathbb{P}$ and  $p_{s_i}=p_2,\forall i \in \mathbb{N}$. Further, let the market in which Agent $i$ contributes be $\Tilde{p}_{s_i}$.

    \subsubsection{Agent Utility}
    The utility for Agent $i \in \mathbb{A}$ with $\Tilde{p}_{s_i}=p_1$, in PPSN is as follows,
   \begin{equation}\tag{PPSN-U1}
	\label{eqn1::PPSN}
    u_i(\cdot) = I_{\mathcal{X}^1\geq h^1}\cdot (\theta_i - x_i) + I_{\mathcal{X}^1< h^1}\cdot \left(R_i - x_i\right)
	\end{equation}
    The utility for Agent $i \in \mathbb{A}$ with $\Tilde{p}_{s_i}=p_2$, in PPSN is as follows,
    \begin{equation}\tag{PPSN-U2}
     u_i(\cdot) = I_{\mathcal{X}^2\geq h^2}\cdot \left(- x_i\right)  + I_{\mathcal{X}^2< h^2}\cdot (\theta_i + R_i - x_i)
	\label{eqn2::PPSN}
	\end{equation}
    
    \subsubsection{Equilibrium Analysis}
	We now provide the equilibrium analysis of this mechanism as the following theorem,

                \begin{theorem}
        \label{Theorem:PPSN}
    For PPSN, with the utility as given by Eq. \ref{eqn1::PPSN} and Eq. \ref{eqn2::PPSN} $\forall i \in \mathbb{A}$, $C:\mathbf{R}^2\rightarrow\mathbf{R}$ as the cost function, $C_0^1:\mathbf{R}\rightarrow\mathbf{R}$ as the cost function obtained from $C$ by fixing the number positive outcome securities satisfying \cite[CONDITION 7]{PPS2016} and used in the market $p_1$ satisfying $(C_0^1)^{-1}(h^1+C_0(0))<\vartheta^1$, and $C_0^2:\mathbf{R}\rightarrow\mathbf{R}$ as the cost function obtained from $C$ by fixing the number of negative outcome securities satisfying\footnote{It is trivial to see that both the cost functions shall be the same. Hence, from hereon we will refer to both of them without the superscript i.e., as $C_0$.} \cite[CONDITION 7]{PPS2016} and used in the market $p_2$ satisfying $C_0^{-1}(h^2+C_0(0))<\vartheta^2$, with $\vartheta^1>h^1$ and $\vartheta^2>h^2,$ a set of strategies in the set $\sigma^*_i=\left(x_i^*,t_i^*,\Tilde{p}_{s_i}\right)$,
     \[
 	  \sigma_i^* = 
 	  \begin{cases}
 	     	(0,a_i,p_{s_i}) \quad \mbox{~if~} \exists \ l \mbox{~s.t.~} h^l_{a_i}=0, \mbox{~else,}\\ 
 	     (x_i^*,a_i,p_{s_i}) : x_i^* \leq C_0(|\theta_i|+Q^{a_i})-C_0(Q^{a_i})
	  \end{cases}
	\]
      $\forall l\in\{1,2\}$; are sub-game perfect equilibria $\forall i \in \mathbb{A}$, such that at equilibrium either $\mathcal{X}^1 = h^1 \mbox{~or~} \mathcal{X}^2 = h^2$ holds.
        \end{theorem}
\noindent\textbf{Proof.} In Step 1, we show that the equilibrium contributions are such that at equilibrium either $\mathcal{X}^1=h^1$ or $\mathcal{X}^2=h^2$ holds. We prove that every agent contributes based on its true preference in Step 2. Step 3 calculates the equilibrium contribution of every agent. In Step 4 we give the conditions for existence of Nash Equilibrium. We show that these set of strategies are sub-game perfect in Step 5. \\
\noindent\underline{\emph{Step 1}}: As $\vartheta^1> h^1$ and  $\vartheta^2> h^2$, at equilibrium $\mathcal{X}^1<h^1$ and $\mathcal{X}^2<h^2$ cannot hold, as $\exists i\in \mathbb{P}\mbox{~and~}\exists j\in \mathbb{N}$ with $x_i<\theta_i$ and $x_j<|\theta_j|$, at least, that could obtain a higher refund bonus by marginally increasing its contribution since $\frac{\partial R_i^{t_i}}{\partial x_i}>1$ \cite[CONDITION 7]{PPS2016}. Likewise, any agent with a positive contribution could gain in utility by marginally decreasing its contribution if $\mathcal{X}^1>h^1$ or $\mathcal{X}^2>h^2$. Thus, at equilibrium, either $\mathcal{X}^1=h^1$ or $\mathcal{X}^2=h^2$ holds. \\
\noindent\underline{\emph{Step 2}}: Since every Agent $i$ is symmetric in its belief towards the project's provision, its expected utility is given by $1/2(\theta_i+R_i)-x_i$ for both the markets. Thus, every Agent $i$ has no incentive to deviate from its preference. Therefore, $\Tilde{p}_{s_i}=p_{s_i},\forall i \in \mathbb{A}$. \\
\noindent\underline{\emph{Step 3}}: As the refund scheme is decreasing with respect to time $t$ (Proposition 2), Agent $i$ contributes as soon as it arrives i.e., at time $a_i$. \\
\noindent\underline{\emph{Step 4}}: Let $q_{p_1}^{a_i}$ be the number of total issued securities at market $p_1$ at time $a_i$, and $q_{p_2}^{a_i}$ be the number of total issued securities at market $p_2$ at time $a_i$, with $Q^{a_i}=\textnormal{min}(q^{a_i}_{p_1},q^{a_i}_{p_2})$ for Agent $i$. Now,
    	\begin{enumerate}
	\item \emph{For Positive agents:} At equilibrium, the \emph{best response} for an Agent $i \in \mathbb{P}$ is that contribution $x_i^*$ in market $p_1$ at time $a_i$ such that its \emph{provisioned utility is not less than not provisioned utility} since it prefers the project to be provisioned and is symmetric in its belief i.e.,
    			$$\theta_i - x_i^* \geq R_i^* - x_i^*$$
                $$\theta_i\geq R_i^*$$
            	\begin{equation*}
            \Rightarrow	x_i^*\leq C_0(\theta_i+Q^{a_i})-C_0(Q^{a_i}) \ \forall i \in \mathbb{P}
            	\end{equation*}
            The result follows from \cite[Eq. 6]{PPS2016}. Based on this $x_i^*$, the number of issued securities changes by $r_{i_{p_1}}^*=C_0^{-1}(x_i^*+C_0(q^{a_i}_{p_1}))-q^{a_i}_{p_1}$ in $p_1$, since a positive agent always contributes in $p_1$. 
    \item \emph{For Negative agents:} At equilibrium, the \emph{best response} for an Agent $i\in \mathbb{N}$ is that contribution $x_i^*$ at time $a_i$ in market $p_2$ such that its \emph{not provisioned utility is not less than provisioned utility} since it prefers the project to not be provisioned and is symmetric in its belief i.e.,
    	$$R_i^* - x_i^* + |\theta_i| \leq -x_i^*$$
    $$R_i^*\leq |\theta_i|$$
            	\begin{equation*}
             \Rightarrow   x_i^*\leq C_0(|\theta_i|+Q^{a_i})-C_0(Q^{a_i}) \ \forall i \in \mathbb{N}
            	\end{equation*}
        The result follows from \cite[Eq. 6]{PPS2016}. Based on this $x_i^*$, the number of issued securities changes by $r_{i_{p_2}}^*=C_0^{-1}(x_i^*+C_0(q^{a_i}_{p_2}))-q^{a_i}_{p_2}$ in $p_2$, since a negative agent always contributes in $p_2$. 
    	\end{enumerate}
\noindent\underline{\emph{Step 4}}: From Proposition 1, $\theta_i\geq R_i^*$ can be written as $\theta_i\geq R_i^* \geq r_{i_{p_1}}^* $; or $r_{i_{p_1}}^*\leq\theta_i \ \forall i \in \mathbb{P}$. Summing up  $\forall i \in \mathbb{P}$, we get the condition for existence of Nash Equilibrium here as, from \cite[Eq. 7]{PPS2016},     
    \begin{equation*}C_0^{-1}(h^1+C_0(0))<\vartheta^1.\end{equation*}
    Similarly for $p_2$, $\forall i \in \mathbb{N}$ we have, 
    \begin{equation*}C_0^{-1}(h^2+C_0(0))<\vartheta^2.\end{equation*}
\noindent\underline{\emph{Step 5}}: For Agent $j$ entering last, if $\mathcal{X}^1=h^1$ or $\mathcal{X}^2=h^2$, then its best response is contributing $0$. If $\mathcal{X}^1<h^1$ and $\mathcal{X}^2<h^2$, irrespective of the total contribution, its provisioned and not provisioned utility is the same at $x_j^*$, defined in the theorem, and it is best response for Agent $j$ to follow the equilibrium strategy. With backward induction, by similar reasoning, it is best response for every agent to follow the equilibrium strategy irrespective of the history of the contributions. 

For Agent $j\in \mathbb{P} (j\in \mathbb{P})$ entering the market such that $h^{1}_{a_j}<x_j^*(h^{2}_{a_j}<x_j^*)$, its best response will be contributing $h^1_{a_j} (h^2_{a_j})$. This is because for the contribution $h^{1}_{a_j}<x_j^*(h^{2}_{a_j}<x_j^*)$, its provisioned utility will be greater than its not provisioned utility. Agent $j$ will also contribute the maximum contribution, $h^1_{a_j} (h^2_{a_j})$, since its not provisioned utility increases as its contribution increases. Therefore, contributing an amount less than $h^1_{a_j} (h^2_{a_j})$ will result in a lesser not provisioned utility for the agent.  Thus, these strategies form a set of sub-game perfect equilibria. \qed\\
    	
\noindent\underline{\textbf{Discussion:}} For PPSN, it can be seen that $\vartheta = \vartheta^1-\vartheta^2$. The project is always provisioned if $\vartheta^1 > h^1$ and $\vartheta \geq 0$ or is never provisioned if $\vartheta^2 > h^2$ and $\vartheta < 0$. Here it must be noted that it can happen that $\vartheta^1 > h^1$ and $\vartheta^2 > h^2$ are simultaneously satisfied. In that case, if $\vartheta^1 > \vartheta^2$, the project attains provision point faster than rejection point and vice-versa. 

The significance of this result is that, in PPSN (and PPRN), at equilibrium, the project is provisioned if the majority \emph{prefers} it, i.e., only when $\vartheta\geq 0$. Thus, this methodology allows for truthful aggregation of private preferences of each agent with respect to public projects.


\section{Civic Crowdfunding for Agents with Asymmetric Beliefs}
\label{section::assymmetric}
In this section, we present a \emph{General Mechanism} which incentivizes agents with asymmetric beliefs towards the public project's provision, to contribute towards it. In this section, we \emph{restrict} our attention to the case where every agent has a \emph{positive} valuation towards project's provision.

The General Mechanism involves two phases: a \emph{Belief Phase} (BP) and a \emph{Contribution Phase} (CP). In BP, each Agent $i$ submits its belief for the provision of the project for which it is allocated some share (denoted by $b_i$) of the reward calculated through Belief Based Reward (BBR) scheme described in the next subsection. In CP, each Agent $i$ submits its contribution ($x_i$) to the project which is dependent on the refund obtained in the BP as well as on the provision point mechanism deployed for civic crowdfunding.

The mechanism requires two separate bonuses for both the phases, which the PM announces at the start of the project. Let $B^B (B^C)$ be the bonus allocated for the BP and the CP, respectively. Further, let $a_i^1(a_i^2)$  be the time at which Agent $i$ arrives to the mechanism for the BP (CP) with $t_i^1 (t_i^2)$ as the time at which it reports its belief (contribution). Let the deadline for the BP (CP) be $T^B (T^C)$  announced at the start of the project.

Unlike in the case of civic crowdfunding for agents with symmetric beliefs, an asymmetric agent which has significant belief towards the project getting provisioned or not, may choose to free-ride and not contribute. Therefore, we introduce a reward scheme that further incentivizes such agent's to contribute towards the project. 

\subsection{Belief Based Reward (BBR)}
To quantitatively measure the reward share to be distributed to every contributing agent in the BP,  we use a PPM, $\mathcal{M}$. We consider PPMs which incentivize truthful elicitation of an agent's belief i.e., PPMs which are IC.

Let the score of Agent $i$ dependent on its belief $(1/2\pm \epsilon_i)$ be $\mathbb{M}_i$. Further, let $S^{t_i}$ be the set consisting of all the agents that have reported their belief including the Agent $i$, who reports its belief at time $t_i$. For $T^B$ as the deadline, $S^{T^B}$ consists of all the agents that have reported their belief. Let $\mathbb{M}_i,\forall i \in S^{T^B}$ be the agent scores calculated after the deadline. For, $$w_i = \frac{\mathbb{M}_i}{\sum_{j} \mathbb{M}_j} \ \forall j \in S^{t_i},$$ Agent $i$'s reward in the scheme is, 
     \begin{equation}
     \label{eqn::BBR}
 	  b_i = 
 	  \begin{cases}
 	    \frac{w_i}{\sum_j w_j} \times B^B \quad \forall j \in \mathbb{A}^+ ;  & \forall i \in \mathbb{A}^+    \\
        \frac{w_i}{\sum_j w_j} \times B^B \quad \forall j \in \mathbb{A}^-;  & \forall i \in \mathbb{A}^-  \\
	  \end{cases}
	\end{equation}
   
 	We refer to the reward scheme given by Eq. \ref{eqn::BBR} as Belief Based Reward (BBR). With this, we show the following proposition:
        \begin{proposition}
        \label{propo::BBR}
        BBR is a decreasing function of time.
        \end{proposition}
        \noindent\textbf{Proof.} From Eq. \ref{eqn::BBR}, BBR is inversely proportional to the order in which agents report their beliefs. Since the arrival of agents is non-decreasing w.r.t. time, BBR is a decreasing function of time. \qed

	In addition, BBR is also \emph{strongly budget balanced}, i.e., in BBR the entire budget is utilized. Note that, at the end of the mechanism, only one set of agents, either $\mathbb{A}^+$ or $\mathbb{A}^-$, are rewarded.\\

\noindent\underline{\textbf{RBTS Reward Scheme:}} In this paper, we use RBTS Mechanism to calculate the mechanism score $\mathbb{M}_i$ for each Agent $i$. For this, every agent submits its prediction and information report as described earlier.

In this reward scheme, let $f_i=0$ denote that Agent $i$ has belief that the project will be provisioned and $f_i=1$ denote that Agent $i$ has belief that the project will not be provisioned. Thus, through each agent's prediction report, the PM knows whether an agent belongs to the set $\mathbb{A}^+$ or the set $\mathbb{A}^-$.

We now present \emph{Provision Point Mechanism with Refunds for Agents with Asymmetric Beliefs} (PPRx) by plugging PPR refund bonus scheme for the CP.

\subsection{Provision Point Mechanism with Refunds for Agents with Asymmetric Beliefs (PPRx)}
    
In this mechanism, we plugin PPR refund bonus scheme for the Contribution Phase. 
       
	\subsubsection{Agent Utility} The utility for Agent $i \in \mathbb{A}^+$, in PPRx is as follows,
   \begin{equation}\tag{PPRx-U1}
	\label{eqn1::PPRx}
    u_i(\cdot) = I_{\mathcal{X}\geq h^0}\cdot (\theta_i - x_i + b_i) + I_{\mathcal{X}< h^0}\cdot \left(\left(\frac{x_i}{\mathcal{X}}\right)B^C\right)
	\end{equation}
    Similarly, the utility for Agent $i \in \mathbb{A}^-$, in PPRx are as follows,
    \begin{equation}\tag{PPRx-U2}
	\label{eqn2::PPRx}
        u_i(\cdot)= I_{\mathcal{X}\geq h^0}\cdot ( \theta_i - x_i) + I_{\mathcal{X}< h^0}\cdot \left(\left(\frac{x_i}{\mathcal{X}}\right)B^C + b_i\right)
	\end{equation}
   
   \subsubsection{Equilibrium Analysis}
   We present the equilibrium analysis of PPRx as the following theorem,
    \begin{theorem}
    \label{Theorem::PPRx}
			For PPRx, with the utility as given by Eq. \ref{eqn1::PPRx} and Eq. \ref{eqn2::PPRx} $\forall i \in \mathbb{A}$, $\vartheta+B^B \geq h^0$ with $B^B,B^C>0$ and Belief Phase reward calculated as per Eq. 3 $\forall i\in \mathbb{A}$, a set of strategies $\sigma_i^* = (x_i^*,t_i^{1*},t_i^{2*})$, 
     \[
 	  \sigma_i^* = 
 	  \begin{cases}
      	(0,a^1_i,T^{C}) \ \mbox{~if~} h^0_{T^C} = 0, else,\\
        (x_i^*,a^1_i,T^{C}) : x_i^* \leq  \left(\frac{k_i^1\theta_i+k_i^1b_i}{k^2_iB^C+k^1_ih^0}\right)h^0 & otherwise
	  \end{cases}
	\] $\forall i \in \mathbb{A}^+$ and a set of strategies $\sigma_i^* =(x_i^*,t_i^{1*},t_i^{2*})$,
    \[
 	  \sigma_i^* = 
 	  \begin{cases}
      	(0,a^1_i,T^{C}) \ \mbox{~if~} h^0_{T^C} = 0, else,\\
        (x_i^*,a^1_i,T^{C}) : x_i^* \leq \left(\frac{k_i^2\theta_i-k_i^1b_i}{k^1_iB^C+k^2_ih^0}\right)h^0 & otherwise
	  \end{cases}
	\] $\forall i \in \mathbb{A}^-$; form a set of equilibrium strategies for the respective set of agents, such that at equilibrium $\mathcal{X} = h^0$ holds.   		
   		\end{theorem}
        \noindent\textbf{Proof.}
 In Step 1, we show that the equilibrium contributions are such that at equilibrium $\mathcal{X}=h^0$ holds. We prove that every agent reports its belief as soon as it arrives to the Belief Phase in Step 2. Step 3 calculates the equilibrium contributions of every agent. Finally, in Step 4 we give the conditions for existence of Nash Equilibrium. \\
\noindent\underline{\emph{Step 1}}: At equilibrium, $\mathcal{X}=h^0$, since the PM stops the protocol as soon as the provision point is reached. Further if $\mathcal{X}<h^0$, then an agent can increase its utility by contributing more to the project, and receiving a higher utility since $B^C > 0$. Therefore, the contributions are such that the market is provisioned at equilibrium.\\
\noindent\underline{\emph{Step 2}}: Since the reward scheme for the Belief Phase is a decreasing function time (Proposition 3), a rational Agent $i$ would report its belief as soon as it arrives to the phase i.e., at time $a_i^1$. \\
\noindent\underline{\emph{Step 3}}: The equilibrium strategy for each agent $i \in \mathbb{A}$ is such that its \emph{provisioned utility is not less than its not provisioned utility}. Now,
		\begin{enumerate}
		\item For agent $i \in \mathbb{A}^+$: 
        	$$ k_i^1(\theta_i - x_i + b_i) \geq k_i^2\left(\frac{x_i}{\mathcal{X}}B^C\right)$$
            $$ \Rightarrow  x_i^* \leq  \left(\frac{k_i^1\theta_i+k_i^1b_i}{k^2_iB^C+k^1_ih^0}\right)h^0.$$
        Since at equilibrium $\mathcal{X}=h^0$.
        \item For agent $i \in \mathbb{A}^-$: 
        	$$ k_i^2(\theta_i - x_i) \geq k_i^1\left(\frac{x_i}{\mathcal{X}}B^C + b_i\right)$$
            $$ \Rightarrow   x_i^* \leq \left(\frac{k_i^2\theta_i-k_i^1b_i}{k^1_iB^C+k^2_ih^0}\right)h^0. $$ 
        Since at equilibrium $\mathcal{X}=h^0$.
		\end{enumerate}
\noindent\underline{\emph{Step 4}}:Note that from Eq. \ref{eqn1::PPRx}, $x_i^*\leq \theta_i + b_i\ \forall i \in \mathbb{A}^+$ and from Eq. \ref{eqn2::PPRx}, $x_i^*\leq \theta_i \ \forall i \in \mathbb{A}^-$. Thus, $$\sum_{i\in \mathbb{A}} x_i^* \leq \sum_{i\in \mathbb{A}^+}(\theta_i+b_i)+\sum_{i\in \mathbb{A}^-}(\theta_i)$$ $$\sum_{i\in \mathbb{A}} x_i^* \leq \vartheta + B^B.$$
        At equilibrium we have $\sum_{i \in \mathbb{A}} x_i^*=h^0$. Therefore,
        $$h^0\leq \vartheta+B^B,$$ is the condition for the existence of Nash Equilibrium. \qed

In the next subsection, we present \emph{Provision Point Mechanism with Securities for Agents with Asymmetric Beliefs} (PPSx) by plugging PPS refund bonus scheme for the CP.

    \subsection{Provision Point Mechanism with Securities for Agents with Asymmetric Beliefs (PPSx)}
	
In this mechanism, we plugin PPS refund bonus scheme for the Contribution Phase. 
    
	\subsubsection{Agent Utility} The utility for Agent $i \in \mathbb{A}^+$, in PPSx is as follows,
   \begin{equation}\tag{PPSx-U1}
	\label{eqn1::PPSx}
    u_i(\cdot) = I_{\mathcal{X}\geq h^0}\cdot (\theta_i - x_i + b_i) + I_{\mathcal{X}< h^0}\cdot \left(r_i - x_i \right)
	\end{equation}
    Similarly, the utility for Agent $i \in \mathbb{A}^-$, in PPSx is as follows,
    \begin{equation}\tag{PPSx-U2}
	\label{eqn2::PPSx}
        u_i(\cdot) = I_{\mathcal{X}\geq h^0}\cdot (\theta_i - x_i) + I_{\mathcal{X}< h^0}\cdot \left(r_i - x_i + b_i \right)
	\end{equation}

	\subsubsection{Equilibrium Analysis}
    We present the equilibrium analysis of PPSx as the following theorem,
    	\begin{theorem}
    	\label{Theorem::PPSx}
For PPSx, with the utility as given by Eq. \ref{eqn1::PPSx} and Eq. \ref{eqn2::PPSx}, $\forall i \in \mathbb{A}^+$, $C:\mathbf{R}^2\rightarrow\mathbf{R}$ as the cost function, with $C_0:\mathbf{R}\rightarrow\mathbf{R}$ as the cost function obtained from $C$ by fixing the number positive outcome securities satisfying \cite[CONDITION 7]{PPS2016}, $\vartheta+B^B \geq h^0$ with $B^B,B^C>0$ and Belief Phase reward calculated as per Eq. \ref{eqn::BBR} $\forall i\in \mathbb{A}$, a set of strategies $\sigma_i^* =(x_i^*,t_i^{1*},t_i^{2*})$ in the set,
	\[
 	  \sigma_i^* = 
 	  \begin{cases}
      	(0,a_i^1,a_i^2); \ \mbox{~if~} h^0_{a_i^2} = 0, \mbox{~else},\\
        (x_i^*,a_i^1,a_i^2) : x_i^* \leq C_0(\theta_i+b_i+q^{a_i^2})-C_0(q^{a_i^2}),
	  \end{cases}
	\]
 $\forall i \in \mathbb{A}^+$ and a set of strategies $\sigma_i^* =(x_i^*,t_i^{1*},t_i^{2*})$ in the set,
\[
 	  \sigma_i^* = 
 	  \begin{cases}
      	(0,a_i^1,a_i^2); \ \mbox{~if~} h^0_{a_i^2} = 0, \mbox{~else},\\
        (x_i^*,a_i^1,a_i^2) : x_i^* \leq C_0(\theta_i-b_i+q^{a_i^2})-C_0(q^{a_i^2}),
	  \end{cases}
	\]
 $\forall i \in \mathbb{A}^-$; are sub-game perfect equilibria such that at equilibrium $\mathcal{X}=h^0$ holds.
    	\end{theorem}
\noindent \textbf{Proof.} In Step 1, we show that the equilibrium contributions are such that at equilibrium $\mathcal{X}=h^0$ holds. We prove that every agent reports its belief as soon as it arrives to the Belief Phase as well as contributes as soon as it arrives to the Contribution Phase in Step 2. Step 3 calculates the equilibrium contributions of every agent. In Step 4, we give the conditions for existence of Nash Equilibrium. We show that these set of strategies are sub-game perfect in Step 5. \\
\noindent\underline{\emph{Step 1}}: As $\vartheta+B^B\geq h^0$, at equilibrium $\mathcal{X}<h^0$ cannot hold, as $\exists i\in \mathbb{A}^+$ with $x_i<\theta_i+b_i$ or $\exists i\in \mathbb{A}^-$ with $x_i<\theta_i$, at least, that could obtain a higher refund bonus by marginally increasing its contribution, since $\frac{\partial r_i^{t_i}}{\partial x_i}>1$ \cite[CONDITION 7]{PPS2016}. Likewise, any agent with a positive contribution could gain in utility by marginally decreasing its contribution if $\mathcal{X}>h^0$. Thus, at equilibrium $\mathcal{X}=h^0$.\\
\noindent\underline{\emph{Step 2}}: Since the reward scheme for the Belief Phase is a decreasing function of time (Proposition 3), a rational Agent $i$ would report its belief as soon as it arrives to the BP i.e., at time $a_i^1$. Also, since the CP is the PPS mechanism, the best response for any for any agent is also to contribute as soon as it arrives i.e., at time $a^2_i$.\\
\noindent\underline{\emph{Step 3}}: From the utility of Agent $i$ in PPSx (Eq. 4 and Eq. 5), it is clear to see that $\forall i \in \mathbb{A}^+$, $x_i \leq \theta_i+b_i$ (as a strategic agent will not contribute greater than its valuation and the reward it received) and $\forall i \in \mathbb{A}^-$, $x_i \leq \theta_i$. Also, $x_i \geq 0$. Further, the equilibrium strategy for each Agent $i \in \mathbb{A}$ is such that its \emph{provisioned utility is not less than its not provisioned utility}. Now,
   	\begin{enumerate}[leftmargin=*]
   	    \item \emph{For Agent $i\in \mathbb{A}^+$:} 
            $$k^1_i(\theta_i+b_i-x^*_i)\geq k^2_i(r_i^*-x_i^*)$$ $$\Rightarrow x_i^*\leq\frac{k^1_i(\theta_i+b_i)-k^2_i(r_i^*)}{k^1_i-k^2_i}.$$
            
            Thus, the maximum value of $x_i^*$ is,
            $$\hat{x}^{*}_{i}=\frac{k^1_i(\theta_i+b_i)-k^2_i(r_i^*)}{k^1_i-k^2_i}$$
            
            From Eq. 4, we have $x_i\leq \theta_i+b_i,\forall i\in \mathbb{A}^+$. Therefore, the maximum value should also be less than $\theta_i+b_i$, i.e., $$\hat{x}^*_{i}\leq\theta_i+b_i \Rightarrow r_i^*\leq \theta_i+b_i.$$ This follows from the value of $\hat{x}_i^*$ defined above. To obtain $r_i^*$ securities at equilibrium, an Agent $i$'s contribution $x^*_i$ must be $$\Rightarrow x^*_i\leq C_0(\theta_i+b_i+q^{a_i^2})-C_0(q^{a_i^2}) \quad \forall i \in \mathbb{A}^+.$$
            The result follows from \cite[Eq. 6]{PPS2016}, i.e., the securities obtained $(r_i^*)$ are monotonic function of the contribution $(x_i^*)$.
    \item \emph{For Agent $i\in \mathbb{A}^-$:} 
            $$k^1_i(r_i^* - x_i^* + b_i)\leq k^2_i(\theta_i-x_i^*).$$ 
           Similar to (1) of this step, the equilibrium contribution $x_i^*$ becomes $$\Rightarrow x_i^*\leq C_0(\theta_i-b_i+q^{a_i^2})-C_0(q^{a_i^2}) \quad \forall i \in \mathbb{A}^-.$$ This follows from $x_i^* \leq \theta_i, \forall i \in \mathbb{A}^-$ and \cite[Eq. 6]{PPS2016}.
            
            \end{enumerate}
\noindent\underline{\emph{Step 4}}: Note that from Eq. 4, $x_i^*\leq \theta_i + b_i\ \forall i \in \mathbb{A}^+$ and from Eq. 5, $x_i^*\leq \theta_i \ \forall i \in \mathbb{A}^-$. Thus, $$\sum_{i\in \mathbb{A}} x_i^* \leq \sum_{i\in \mathbb{A}^+}(\theta_i+b_i)+\sum_{i\in \mathbb{A}^-}(\theta_i)$$ 
        At equilibrium we have $\sum_{i \in \mathbb{A}} x_i^*=h^0$. Therefore,
        $$h^0\leq \vartheta+B^B,$$ is the condition for the existence of Nash Equilibrium. \\
\noindent\underline{\emph{Step 5}}: For Agent $j$ entering last, if $\mathcal{X}=h^0$, then its best response is contributing $0$. If $\mathcal{X}<h^0$, irrespective of the total contribution, its provisioned and not provisioned utility is the same at $x_j^*$, defined in the theorem, and it is best response for Agent $j$ to follow the equilibrium strategy. With backward induction, by similar reasoning, it is best response for every agent to follow the equilibrium strategy irrespective of the history of the contributions. 

For Agent $j$ entering the market such that $h^0_{a_j^2}<x_j^*$, its best response will be contributing $h^0_{a_j^2}$ as for the contribution $h^0_{a_j^2}<x_j^*$, its provisioned utility will be greater than its not provisioned utility. Agent $j$ will also contribute the maximum contribution, $h^0_{a_j^2}$, since its not provisioned utility increases as its contribution increases. Therefore, contributing an amount less than $h^0_{a_j^2}$ will result in a lesser not provisioned utility for the agent.  Thus, these strategies form a set of sub-game perfect equilibria. \qed

  \subsection{Equilibrium Contribution Analysis} 
    We now compare the equilibrium contribution of both the sets of agents in PPRx and PPSx. Towards this, let Agent $i\in \mathbb{A}^+$ and Agent $j\in\mathbb{A}^-$ such that $\theta_i=\theta_j$, $b_i=b_j$, $a_i^2=a_j^2$ and $k_i^1=k_j^1$ ($k^1_i+k^2_i=1$). Agent $i$'s belief about the project getting provisioned is $k_i^1$ and Agent $j$'s $k_j^2$.
        \begin{enumerate}[leftmargin=*]
            \item \underline{\emph{For PPRx}}: The difference in the equilibrium contribution of Agent $i\in\mathbb{A}^+$ and Agent $j\in\mathbb{A}^-$ as defined above in PPRx now becomes,
        \begin{align*}
        x_i^*-x_j^*&=\left(\frac{k_i^1\theta_i+k_i^1b_i}{k^2_iB^C+k^1_ih^0}\right)h^0-\left(\frac{k_i^2\theta_i-k_i^1b_i}{k^1_iB^C+k^2_ih^0}\right)h^0
        \end{align*}
        As the denominator is always positive, in the RHS of the above equation we only consider the numerator. Observe,
        $$
        \begin{aligned} 
        h_0(k^1_iB^C+k^2_ih^0)(k_i^1\theta_i+k_i^1b_i)-&\\h_0(k^2_iB^C+k^1_ih^0)(k_i^2\theta_i-k_i^1b_i) > 0
        \end{aligned}
        $$
        $$
        \begin{aligned}
        \mbox{as,~} B^C\theta_i((k_i^1)^2-(k_i^2)^2)+B^Cb_i((k_i^1)^2+k_i^1k_i^2)+&\\h^0b_i((k_i^1)^2+k_i^1k_i^2) > 0,
        \end{aligned}
        $$
        since $k_i^1\geq k_i^2$. Thus, the upper bound of each Agent $i\in\mathbb{A}^+$ is always greater than for an Agent $j\in\mathbb{A}^-$ with the same valuation and belief. 
        
        \item \underline{\emph{For PPSx}}: For Agent $i\in\mathbb{A}^+$ and Agent $j\in\mathbb{A}^-$ as defined above, the equilibrium contribution of Agent $i$ will always be greater than that of Agent $j$ since $C_0(\theta_i+b_i+q^{a_i^2})>C_0(\theta_i-b_i+q^{a_i^2})$ as $b_i>0$ and $\frac{\partial r_i^{t_i}}{\partial x_i}>1$ \cite[CONDITION 7]{PPS2016}.
        \end{enumerate}
    
    Thus, for both PPRx and PPSx, the upper bound on the equilibrium contributions $\forall i \in \mathbb{A}^+$ (with $k^1_i$ as the belief towards project's provision) is greater than the upper bound on the equilibrium contributions $\forall i \in \mathbb{A}^-$ (with $k^2_i$ as the belief towards project's provision; $k^1_i+k^2_i=1$), for the same valuation and belief. 
    
    This implies that agents with greater belief towards the project's provision contribute more than agents with lesser belief towards it. Thus, BBR (Eq. \ref{eqn::BBR} and the utility structure as given by Eqs. \ref{eqn1::PPRx}, \ref{eqn2::PPRx} for PPRx and Eqs. \ref{eqn1::PPSx}, \ref{eqn2::PPSx} for PPSx, provides a natural way for civic crowdfunding with asymmetric agents such that the project is provisioned at equilibrium.
    
    In the next section, we discuss the setting up of the markets for all these mechanisms.
    
    \section{Discussion}
    For PPSN (PPRN), the PM is required to set up two independent PPS (PPR) markets. The provision point for these projects are determined based on the economics of their construction. The rejection point can be similarly determined. For instance, the rejection point for our garbage dump yard example could be the cost of constructing the dump yard at a different locality. Another method for determining the rejection point could be the cost incurred by the government as a result of the public project not getting provisioned. An instance of this could be the construction of dams. The cost of not setting up the dam, i.e., the rejection point for the project could be the cost incurred by the government in providing electricity or water etc. to the nearby areas which they could have achieved through the dam's construction. Note that, the amount collected if the project is rejected is at the discretion of the government.

The PM should allocate reasonable budget for all these mechanisms. Allocating huge budgets may not guarantee the provision/rejection of the projects. In such a case, the agents may prefer to contribute just enough to get substantial refunds. Likewise, allocating insignificant budgets may prove to not be incentivising enough for agents to contribute to the market.

In PPSN, the cost function, $C_0$, used to allocate the securities must also be same for both markets. Additional details for setting up the prediction markets as well as the budget can be found at \cite{PPS2016,REPPS2017}.

\subsection{Designing Mechanisms for Asymmetric Agents with Negative Valuation}

Civic crowdfunding for agents with information structure consisting of both -- their preference and their belief towards the provision of the project, is not trivial, as it provides an extra dimension for the agents to manipulate the mechanism. For instance, combining PPSN and PPSx (PPRN and PPRx) will not suffice. An Agent $i\in \mathbb{A}^+$ with $\theta_i\geq 0$, will \emph{always} choose to contribute towards the project not getting provisioned, as it believes that the project will be provisioned anyways, making it eligible for the additional refund bonus. Likewise, an Agent $i\in \mathbb{A}^-$ with $\theta_i<0$ will always contribute towards the provision of the project. However, an Agent  $i\in \mathbb{A}^+$ with $\theta_i<0$ and an Agent $i\in \mathbb{A}^-$ with $\theta_i\geq0$ will always contribute as per their true preference.

For instance, this intuitive result can be shown as follows, from Eq. 1 and Eq. 2 in PPSN with BBR as defined in Eq. 3, for an Agent $i\in \mathbb{A}^+$ with $\theta_i\geq 0$, the difference in its expected utility in contributing in both the markets can be given as,
    $$k_i^1(\theta_i-x_i+b_i)+k_i^2(R_i-x_i)-k_i^1(\theta_i+R_i-x_i+b_i)-k_i^2(-x_i)$$
    $$\Rightarrow (k_i^2-k_i^1)\cdot(R_i) \leq 0,$$
as $\forall i \in\mathbb{A}^+$, $k_i^1\geq k_i^2$ and $R_i\geq0$. Thus, a strategic Agent $i\in \mathbb{A}^+$ with $\theta_i\geq 0$ will always lie about its preference by contributing against the provision of the project. Likewise, an Agent $i\in \mathbb{A}^-$ with $\theta_i<0$ will always contribute towards the provision of the project. However, an Agent  $i\in \mathbb{A}^+$ with $\theta_i<0$ and an Agent $i\in \mathbb{A}^-$ with $\theta_i\geq0$ will always contribute as per their true preference.

Thus, the general method and the general mechanism proposed in this paper for civic crowdfunding for agents with negative valuation and agents with asymmetric belief respectively, are not sufficient to incentivize every asymmetric agent to contribute as per their true preference. This can be further explored in future work.

\section{Conclusion}
In this paper, we explored the limitations of existing literature on civic crowdfunding. We showed that it poses restrictions on the information structure of agents, as it only allows for positive as well as symmetric agents. We broke this barrier on the information structure of an agent by proposing (i) a general methodology for addressing symmetric agents with negative preferences based on which we proposed two mechanisms, PPRN and PPSN (Theorem \ref{Theorem:PPSN}); and (ii) a general mechanism for positive agents with asymmetric beliefs based on which we proposed two mechanisms, PPRx and PPSx (Theorem \ref{Theorem::PPSx}).

We leave it for future work to explore the feasibility of combining negative preferences and asymmetric belief into one framework for civic crowdfunding.
\printbibliography

\end{document}